\documentclass[10pt,journal,a4paper,twoside,twocolumn]{IEEEtran}
\usepackage{cite}
\usepackage{float}
\usepackage{stfloats}
\usepackage{fancyhdr}
\usepackage{color}
\usepackage[inline]{enumitem}
\usepackage{amsmath,amsfonts,amssymb}
\usepackage{graphicx}
\usepackage{algorithm}
\usepackage{algorithmic}
\usepackage{epsfig}
\usepackage{subfigure}
\usepackage{epstopdf}
\usepackage{bm}
\usepackage{array}

\usepackage{bbm}

\newtheorem{Lemma}{Lemma}
\newtheorem{Rem}{Remark}

\begin{document}
%	\pagenumbering{number}
	\title{Low-Complexity Beamforming Design for Null Space-based Simultaneous Wireless Information and Power Transfer Systems
	}
	
	% \author{\IEEEauthorblockN{
	% 		Cheng Luo, \emph{Student Member, IEEE}, Jie Hu, \emph{Senior Member, IEEE}, Luping Xiang, \emph{Member, IEEE} and Kun Yang, \emph{Fellow, IEEE}
    %         }
	% 		\\
    %     \thanks{Cheng Luo, Jie Hu and Luping Xiang are with the School of Information and Communication Engineering, University of Electronic Science and Technology of China, Chengdu, 611731, China, email: chengluo@std.uestc.edu.cn; hujie@uestc.edu.cn; luping.xiang@uestc.edu.cn.}
    %     \thanks{Kun Yang is with the Yangtze Delta Region Institute (Quzhou), University of Electronic Science and Technology of China, Quzhou, 324000, China and also with the School of Computer Science and Electronic Engineering, University of Essex, Colchester, CO4 3SQ, U.K. (e-mail: kunyang@essex.ac.uk).}
	% }

	\author{\IEEEauthorblockN{Cheng Luo$^1$, Jie Hu$^{1}$, Luping Xiang$^{2,*}$ and Kun Yang$^{3}$} \\
		\IEEEauthorblockA{
			$^1$ School of Information and Communication Engineering, University of Electronic Science and Technology of China (UESTC), Chengdu, Sichuan, China. \\
			$^2$ State Key Laboratory of Novel Software Technology, Nanjing University, Nanjing 210008, China, and School of Intelligent Software and Engineering, Nanjing University (Suzhou Campus), Suzhou.\\
			$^3$ School of Computer Science and Electronic Engineering, University of Essex, Colchester CO4 3SQ, U.K..\\
			$^{*}$Corresponding Author, Email: luping.xiang@nju.edu.cn
		}
	}
	\maketitle
	
	\thispagestyle{empty} % IEEE模板在\maketitle 后会自动声明\thispagestyle{plain}，
	% 导致第一页什么都没有。所以得把plain更改为fancy
	% \lhead{} % 页眉左，需要东西的话就在{}内添加
	% \chead{} % 页眉中
	% \rhead{} % 页眉右
	% \lfoot{} % 页眉左
	% \cfoot{} % 页眉中
	% \rfoot{\thepage} %页眉右，\thepage 表示当前页码
	\renewcommand{\headrulewidth}{0pt} %改为0pt 即可去掉页眉下面的横线
	\renewcommand{\footrulewidth}{0pt} %改为0pt 即可去掉页脚上面的横线
	% \pagestyle{fancy}

    % \rfoot{\thepage} % 页眉右

	\begin{abstract}
	Simultaneous wireless information and power transfer (SWIPT) is a promising technology for the upcoming sixth-generation (6G) communication networks, enabling internet of things (IoT) devices and sensors to extend their operational lifetimes. In this paper, we propose a SWIPT scheme by projecting the interference signals from both intra-wireless information transfer (WIT) and inter-wireless energy transfer (WET) into the null space, simplifying the system into a point-to-point WIT and WET problem. Upon further analysis, we confirm that dedicated energy beamforming is unnecessary. In addition, we develop a low-complexity algorithm to solve the problem efficiently, further reducing computational overhead. Numerical results validate our analysis, showing that the computational complexity is reduced by 97.5\% and 99.96\% for the cases of $K^I = K^E = 2$, $M = 4$ and $K^I = K^E = 16$, $M = 64$, respectively.
	\end{abstract}

\section{Introduction}

\IEEEPARstart{I}{n} the forthcoming era of sixth-generation (6G) communication networks, a vast number of internet of things (IoT) devices and sensors will be deployed to support a wide range of applications, including healthcare, environmental monitoring, smart homes, smart cities, autonomous vehicles, national defense, and other high-maintenance environments \cite{clerckx2021wireless, 10463057}. The density of IoT devices is expected to reach tens or more per square meter in future networks \cite{lopez2021massive}. While the proliferation of these devices promises significant societal and economic benefits, their limited battery life poses a challenge, necessitating frequent replacements or manual recharging, which, in turn, escalates maintenance costs.

Wireless power transfer (WPT) has emerged as a promising solution to overcome the battery limitations of IoT devices, garnering substantial attention across various domains, from hardware design to resource allocation strategies \cite{wpt_hardware2allocation, massive_WET}. However, the integration of conventional wireless information transfer (WIT) and WPT within the same radio frequency (RF) band presents notable challenges. This has led to a growing body of research on simultaneous wireless information and power transfer (SWIPT)\cite{SWIPT_hardware,Hujie_modulation,Xujie,wqq_IRSWET,fluid_SWIPT_LX}. Specifically, \cite{SWIPT_hardware} explores a non-linear energy harvester model and proposes a power allocation scheme aimed at maximizing the total harvested power at energy users (EUs), while ensuring minimum signal-to-interference-plus-noise ratios (SINRs) for multiple information users (IUs). In \cite{Hujie_modulation}, the authors provide design guidelines for coding-controlled SWIPT systems, addressing modulation schemes for both single-user and multi-user environments. Additionally, \cite{Xujie} proposes a multi-user multiple-input single-output (MISO) SWIPT system, demonstrating that using at most a single energy beam is optimal for the SWIPT system. Similarly, an intelligent reconfigurable surface (IRS)-assisted SWIPT system is introduced in \cite{wqq_IRSWET}, where the authors prove that transmitting information signals solely from the hybrid access point (HAP) is sufficient to serve both IUs and EUs. 
\begin{figure}
	\centering
	\includegraphics[width=0.8\linewidth]{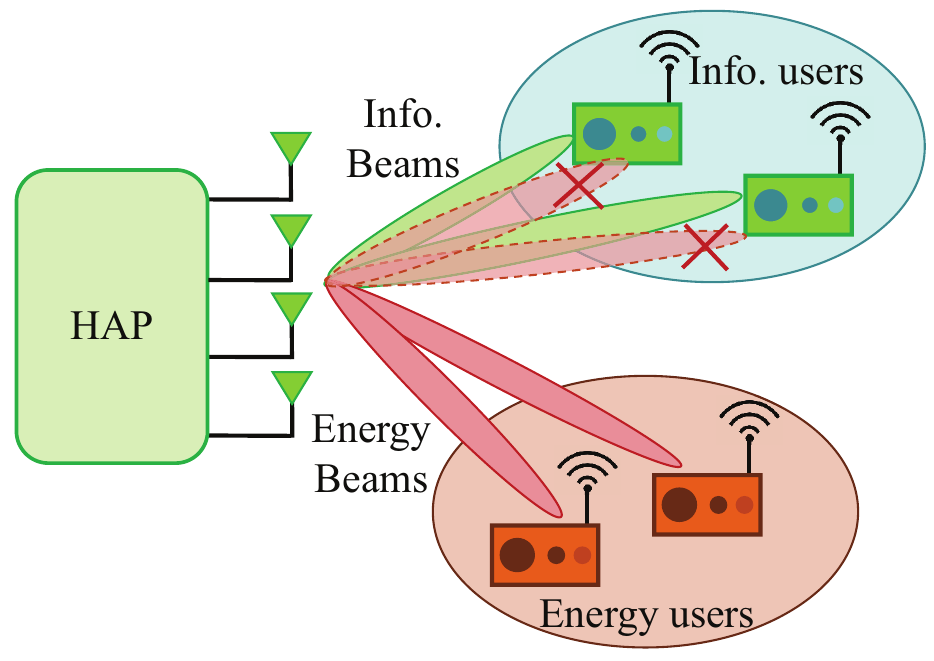}
	\caption{System model of proposed null space-based SWIPT system. The interference signals can be eliminated via null space.}
	\label{fig:systemmodel}
\end{figure}

Moreover, several studies have successfully addressed the interference between WIT and WET by placing interference signals into the null space, achieving excellent performance \cite{nullspace1}. However, the computational complexity of such approaches remains intolerable, particularly in multi-user scenarios. In this paper, we propose a null space-based SWIPT beamforming and its low-complexity design. Our key contributions are as follows:
\begin{itemize}
	\item We propose a null space-based beamforming scheme for SWIPT systems. Upon re-evaluating the design, we conclude that dedicated energy beamforming is unnecessary, which reduces the system's computational complexity.
	\item We develop a low-complexity algorithm to solve the null space-based SWIPT problem, leveraging the fact that the power required for WET is significantly higher than that for WIT. Consequently, this effectively reduces the computational complexity even further.
	\item Numerical results demonstrate the effectiveness of the proposed low-complexity algorithm. Additionally, we observe that a higher Rician factor leads to greater correlation among the channels of users, which negatively impacts the WIT/WET performance due to more severe multiuser interference.
\end{itemize}

The remaining sections of this paper are organized as follows: Section \ref{sec:systemmodel} provides an overview of the system model and problem formulation. Section \ref{sec:nullspacedesign} presents the proposed null space-based SWIPT system and low-complexity design. Numerical results are detailed in Section \ref{sec:numericalresult}, and Section \ref{sec:conclusion} summarizes the findings and conclusions of this paper.

% \emph{Notation:} $\mathbf{I}_{M}$ represents the $M$-dimensional identity matrix. The notation $[\cdot]_i$ and $[\cdot]_{i,j}$ refers to the $i$-th element of a vector and the $(i,j)$-th element of a matrix, respectively. The imaginary unit is denoted by $\mathbbm{i}=\sqrt{-1}$. The Euclidean norm and absolute value are denoted by $||\cdot||$ and $|\cdot|$, respectively. The function $\text{diag}\{\cdot\}$ creates a diagonal matrix. The operators $(\cdot)^{H}$ represents the conjugate transpose. Finally, the notation $\mathcal{CN}$ represents the circularly symmetric complex Gaussian distribution.

\section{System Model}\label{sec:systemmodel}
In this paper, we study a null space-based SWIPT system, As depicted in Fig. \ref{fig:systemmodel}, which involving a HAP with $M$ antennas as a uniform linear array (ULA), $K^E$ single antenna EUs and $K^I$ single antenna IUs. The HAP, providing serves via information beamformings (IBs) and energy beamformings (EBs) simultaneously, and transmitting information signals and energy signals denote by $s^I_i,\forall i\in K^I$ and $s^E_j,\forall j\in K^E$, which are independent and identically distributed (i.i.d) circularly symmetric complex Gaussian (CSCG) random variable with zero mean and unit variance. 
\subsection{Channel model}
Quasi-static block fading Rician channels are taken into consideration. Specifically, the line-of-sight (loS) components of channels for both IUs and EUs are expressed as 
\begin{align}
	\left[\bar{\mathbf{h}}^I_i\right]_m= \frac{1}{\sqrt{M}}e ^{(m-1)\mathbbm{i}\varpi^I_i},\\
	\left[\bar{\mathbf{h}}^E_j\right]_m= \frac{1}{\sqrt{M}}e ^{(m-1)\mathbbm{i}\varpi^E_j},
\end{align}
where $\varpi^I_i = 2\pi d\sin(\theta^{\text{I}}_i)/\lambda=\pi\sin(\theta^{\text{I}}_i), \forall i\in K^I$, $\varpi^E_j = \pi\sin(\theta^{\text{E}}_j), \forall j\in K^E$ by setting $d=\lambda/2$ without loss of generality. $\theta^{\text{I}}_i$ and $\theta^{\text{I}}_i$ denote the angle of departure (AoD) of the $i$-th IU and $j$-th EU, respectively. Thus, the Rician channels of IUs and EUs, denote as $\mathbf{h}^I_i\in\mathbb{C}^{M\times 1}$ and $\mathbf{h}^E_j\in\mathbb{C}^{M\times 1}$, can be expressed as 
\begin{align}
	\mathbf{h}^I_i= \sqrt{\frac{\kappa_i^I}{1+\kappa_i^I}}\bar{\mathbf{h}}^I_i+\sqrt{\frac{1}{1+\kappa_i^I}}\hat{\mathbf{h}}^I_i,\forall i\in K^I,\\
	\mathbf{h}^E_j= \sqrt{\frac{\kappa_j^E}{1+\kappa_j^E}}\bar{\mathbf{h}}^E_j+\sqrt{\frac{1}{1+\kappa_j^E}}\hat{\mathbf{h}}^E_j,\forall j\in K^E,
\end{align}
where $\hat{\mathbf{h}}^I_i\in\mathbb{C}^{M\times 1}$ and $\hat{\mathbf{h}}^E_j\in\mathbb{C}^{M\times 1}$ denote the non-line-of-sight (NLoS) components of corresponding channels. $\kappa^I_i$ and $\kappa^E_j$ denote the Rician factor for channels of $i$-th IU and $j$-th EU, respectively. Thus, we can further express the signal transmitted by the HAP as 
\begin{align}
	\mathbf{x}=\sum_{i\in K^I}\mathbf{w}_i s^I_i +\sum_{j\in K^E}\mathbf{v}_j s^E_j,
\end{align}
where $\mathbf{w}_i,\forall i\in K^I$ and $\mathbf{v}_j,\forall j\in K^E$ denote the $i$-th IB and $j$-th EB, respectively. 

\subsection{Problem formulation}
The signal-to-interference-plus-noise-ratio (SINR) of $i$-th IU can be expressed as 
\begin{align}
	&\eta_i = \nonumber\\
	&\frac{\varrho^I_{i}\left|\left(\mathbf{h}_i^I\mathbf{w}_i\right)^H\right|^2}{\underbrace{\varrho^I_{i}\sum\limits_{l\in K^I, l\neq i}\left|\left(\mathbf{h}_i^I\right)^H\mathbf{w}_l\right|^2+\varrho^I_{i}\sum\limits_{j\in K^E}\left|\left(\mathbf{h}_i^I\right)^H\mathbf{v}_j\right|^2}_{\text{Interference signals}}+\sigma_0^2},\label{eqn:SINR_exp}
\end{align}
where $\varrho^I_{i}$ denotes the pathloss of $i$-th IU and $\sigma_0^2$ denotes the noise power. Moreover, the received energy by the $j$-th EUs can be expressed as 
\begin{align}
	\boldsymbol{\mathcal{E}}_j= \varrho_j^E\sum_{i\in K^I} \left|\mathbf{h}^E_j\mathbf{w}_i\right|^2 + \varrho_j^E\sum_{t\in K^E} \left|\mathbf{h}^E_j\mathbf{v}_t\right|^2,\forall j\in K^E,
\end{align}
where $\varrho^E_{j}$ denotes the pathloss of $j$-th EU. We formulate the objectives for both IUs and EUs as 
\begin{align}
	\text{(P1): }&\max_{\mathbf{w}_i,\forall i\in K^I,\mathbf{v}_j,\forall j\in K^E} \sum_{t\in K^E}\left(\sum_{i\in K^I}\varrho^E_t\left|\left(\mathbf{h}^E_t\right)^H\mathbf{w}_i\right|^2\right.\nonumber\\
	&\qquad\qquad\qquad\qquad\quad\left.+\sum_{j\in K^E}\varrho^E_t\left|\left(\mathbf{h}^E_t\right)^H\mathbf{v}_j\right|^2\right)\label{eqn:obj_o}\\
	&\qquad\quad \text{s.t. } \log_2\left(1+\eta_i\right)\geq \mathcal{T}_i,\forall i\in K^I \tag{\ref{eqn:obj_o}a}\label{eqn:obj_oCa}\\
	&\qquad\qquad \sum_{i\in K^I}||\mathbf{w}_i||^2+\sum_{j\in K^E}||\mathbf{v}_j||^2\leq P_\text{max},\tag{\ref{eqn:obj_o}b}\label{eqn:obj_oCb}
\end{align}
where $\mathcal{T}_i$ denotes the channel capacity required for the $i$-th IU. The objective $\text{(P1)}$ is to maximize the total received power, subject to the communication and transmit power constraints given by \eqref{eqn:obj_oCa} and \eqref{eqn:obj_oCb}. By applying the technique of semidefinite relaxation (SDR), $\text{(P1)}$ can be further expressed as
\begin{align}
	&\text{(P1.1): }\max_{\substack{\mathbf{w}_i,\forall i\in K^I,\\\mathbf{v}_j,\forall j\in K^E}} \quad \sum_{i\in K^I}\mathrm{tr}(\mathbf{S}\mathbf{W}_i)+\sum_{j\in K^E}\mathrm{tr}(\mathbf{S}\mathbf{V}_j)\label{eqn:trace_o}\\
	&\text{s.t. } \frac{\mathrm{tr}\left(\mathbf{h}_i^I\left(\mathbf{h}_i^I\right)^H\mathbf{W}_i\right)}{\left(2^{\mathcal{T}_i}-1\right)}-\sum_{j\in K^E}\mathrm{tr}\left(\mathbf{h}_i^I\left(\mathbf{h}_i^I\right)^H\mathbf{V}_j\right)-\frac{\sigma_0^2}{\varrho^I_i}\nonumber\\
	&\qquad-\sum_{l\in K^I,l\neq i}\mathrm{tr}\left(\mathbf{h}_i^I\left(\mathbf{h}_i^I\right)^H\mathbf{W}_l\right)\geq 0,\forall i\in K^I,\tag{\ref{eqn:trace_o}a}\label{eqn:trace_oCa}\\
	&\,\quad \sum_{i\in K^I}\mathrm{tr}\left(\mathbf{W}_i\right)+\sum_{j\in K^E}\mathrm{tr}\left(\mathbf{V}_j\right)\leq P_\text{max},\tag{\ref{eqn:trace_o}b}\label{eqn:trace_oCb}\\
	&\,\,\quad \mathbf{W}_i\succeq \mathbf{0},\forall i\in K^I, \mathbf{V}_j \succeq\mathbf{0},\forall j\in K^E,\tag{\ref{eqn:trace_o}c}\label{eqn:trace_oCc}
\end{align}
where $\mathbf{S}=\sum_{j\in K^E}\mathbf{h}^E_j\left(\mathbf{h}^E_j\right)^H\in\mathbb{C}^{M\times M}$. $\mathbf{W}_i=\mathbf{w}_i\left(\mathbf{w}_i\right)^H\in\mathbb{C}^{M\times M},\forall i\in K^I$ and $\mathbf{V}_j=\mathbf{v}_j\left(\mathbf{v}_j\right)^H\in\mathbb{C}^{M\times M}$. Eq. \eqref{eqn:trace_o} and constraint are come from $\mathrm{tr}\left(\mathbf{A}\right)+\mathrm{tr}\left(\mathbf{B}\right)=\mathrm{tr}\left(\mathbf{A+B}\right)$. Previous studies have shown that at most one energy beam (EB) is required for energy users (EUs), as detailed in \cite{wqq_IRSWET, Xujie}. Thus, the problem $\text{(P1.1)}$ can be further simplified as
\begin{align}
	&\text{(P1.2): }\max_{\mathbf{w}_i,\forall i\in K^I,\mathbf{v}} \sum_{i\in K^I}\mathrm{tr}(\mathbf{S}\mathbf{W}_i)+\mathrm{tr}(\mathbf{S}\mathbf{V})\label{eqn:trace_simp}\\
	&\text{s.t. } \frac{\mathrm{tr}\left(\mathbf{h}_i^I\left(\mathbf{h}_i^I\right)^H\mathbf{W}_i\right)}{\left(2^{\mathcal{T}_i}-1\right)}-\mathrm{tr}\left(\mathbf{h}_i^I\left(\mathbf{h}_i^I\right)^H\mathbf{V}\right)-\frac{\sigma_0^2}{\varrho^I_i}\nonumber\\
	&\qquad-\sum_{l\in K^I,l\neq i}\mathrm{tr}\left(\mathbf{h}_i^I\left(\mathbf{h}_i^I\right)^H\mathbf{W}_l\right)\geq 0,\forall i\in K^I,\tag{\ref{eqn:trace_simp}a}\label{eqn:trace_simpCa}\\
	&\,\quad \sum_{i\in K^I}\mathrm{tr}\left(\mathbf{W}_i\right)+\mathrm{tr}\left(\mathbf{V}\right)\leq P_\text{max},\tag{\ref{eqn:trace_simp}b}\label{eqn:trace_simpCb}\\
	&\,\,\quad \mathbf{W}_i\succeq \mathbf{0},\forall i\in K^I, \mathbf{V} \succeq\mathbf{0}.\tag{\ref{eqn:trace_simp}c}\label{eqn:trace_simpCc}
\end{align}

Problem $\text{(P1.2)}$ is a convex optimization problem and can be solved by standard optimization method, such as interior point method. The computational complexity of solving $\text{(P1.2)}$ by interior point method is $\mathcal{O}\left(\left((K^I + 1) M\right)^{3.5}\right)$. To obtain the final $\mathbf{w}_i, \forall i\in K^I$ and $\mathbf{v}$, $K^I+1$ times eigenvalue decompositions are required, each with a complexity of $\mathcal{O}\left(\left(K^I+1\right)M^{3}\right)$. Thus, the total computational complexity is $\mathcal{O}\left(\left((K^I + 1) M\right)^{3.5}+\left(K^I+1\right)M^{3}\right)$.

\section{Null Space-Based SWIPT}\label{sec:nullspacedesign}
In this section, we propose a null space-based algorithm to eliminate the interference as shown in Eq. \eqref{eqn:SINR_exp}, while maximize the total received energy. Then, we propose a low computational complexity algorithm to effectively solve the above problem. 
\subsection{Null Space-based Solution}
To eliminate the interference signal expressed in Eq. \eqref{eqn:SINR_exp}, the IBs and EBs need to satisfy 
\begin{align}
	&\left(\mathbf{h}^I_i\right)^H\mathbf{w}_l=0,\forall i\in K^I, l\in K^I, l\neq i,\\
	&\left(\mathbf{h}^I_i\right)^H\mathbf{v}_t=0,\forall i\in K^I, t\in K^E.
\end{align}
Let $\mathbf{H}^I$ and $\mathbf{H}^I_{/i}$ denote the channel matrix involving all the IUs and all IUs except the $i$-th IU, respectively. Specifically, $\mathbf{H}^I=\left[\mathbf{h}^I_1,\cdots,\mathbf{h}^I_{K^I}\right]$ and $\mathbf{H}^I_{/i}=\left[\mathbf{h}^I_1,\cdots,\mathbf{h}^I_{i-1}, \mathbf{h}^I_{i+1},\mathbf{h}^I_{K^I}\right]$. Thus, we can construct the null space as 
\begin{align}
	&\mathbf{N}_i^I=\text{null}\left(\mathbf{H}^I_{/i}\right)\in\mathbb{C}^{M\times \left(M-K^I+1\right)},\label{eqn:NI}\\
	&\mathbf{N}_j^E=\mathbf{N}^E=\text{null}\left(\mathbf{H}^I\right)\in\mathbb{C}^{M\times \left(M-K^I\right)},\label{eqn:NE}
\end{align}
where Eq. \eqref{eqn:NE} holds because all the EUs share the same null space. In practically, we can obtain the null space via singular value decomposition (SVD). Thus, the IBs and EBs can be rewritten as 
\begin{align}
	&\mathbf{w}_i^{\text{null}}=\sqrt{P_i^I}\mathbf{N}_i^I\mathbf{b}_i,\forall i\in K^I,\label{eqn:w_null}\\
	&\mathbf{v}_j^{\text{null}}=\sqrt{P_j^E}\mathbf{N}^E\mathbf{d}_j,\forall j\in K^E,\label{eqn:v_null}
\end{align}
where $\mathbf{b}_i\in\mathbb{C}^{M-K^I+1},\forall i\in K^I$, $\mathbf{d}_j\in\mathbb{C}^{M-K^I},\forall j\in K^E$ denote the normalized vectors. Thus, the problem $\text{(P1)}$ can be expressed as 
\begin{align}
	&\text{(P2): }\max_{\substack{\mathbf{b}_i,\forall i\in K^I,\\\mathbf{d}_j,\forall j\in K^E}} \quad \sum_{t\in K^E}\varrho^E_t\left(\sum_{i\in K^I}\left|\left(\mathbf{h}^E_t\right)^H\sqrt{P_i^I}\mathbf{N}_i^I\mathbf{b}_i\right|^2\right.\nonumber\\
	&\qquad\qquad\qquad\qquad\left.+\sum_{j\in K^E}\left|\left(\mathbf{h}^E_t\right)^H\sqrt{P_j^E}\mathbf{N}^E\mathbf{d}_j\right|^2\right)\label{eqn:obj_null_1}\\
	&\text{s.t. } \log_2\left(1+\frac{\varrho^I_{i}\left|\left(\mathbf{h}_i^I\sqrt{P_i^I}\mathbf{N}_i^I\mathbf{b}_i\right)^H\right|^2}{\sigma_0^2}\right)\geq \mathcal{T}_i,\forall i\in K^I \tag{\ref{eqn:obj_null_1}a}\label{eqn:obj_null_1Ca}\\
	&\,\,\quad \sum_{i\in K^I}P_i^I+\sum_{j\in K^E}P_j^E\leq P_\text{max},\tag{\ref{eqn:obj_null_1}b}\label{eqn:obj_null_1Cb}
\end{align}
where Eq. \eqref{eqn:obj_null_1} is derived by substituting Eq. \eqref{eqn:w_null}-\eqref{eqn:v_null} into Eq. \eqref{eqn:obj_o}, while constraint \eqref{eqn:obj_null_1Ca} arises from the interference elimination. 

Let $\left(\mathbf{h}^{E,I}_{t,i}\right)^H=\left(\mathbf{h}^E_t\right)^H\mathbf{N}_i^I$, $\left(\mathbf{h}^{E}_{t}\right)^H=\left(\mathbf{h}^E_t\right)^H\mathbf{N}^E$, $\left(\mathbf{h}^{I,I}_i\right)^H=\left(\mathbf{h}^{I}_i\right)^H\mathbf{N}^I_i$. Thus, the problem $\text{(P2)}$ can be simplified via SDR as 
\begin{align}
	&\text{(P2.1): }\max_{\substack{\mathbf{b}_i,\forall i\in K^I,\\\mathbf{d},\forall j\in K^E}} \quad \sum_{i\in K^I}\mathrm{tr}(\mathbf{S}^{E,I}_{i}\mathbf{B}_i)+\mathrm{tr}(\mathbf{S}^E\mathbf{D})\label{eqn:trace_simp_null}\\
	&\text{s.t. } \frac{\mathrm{tr}\left(\mathbf{h}_i^{I,I}\left(\mathbf{h}_i^{I, I}\right)^H\mathbf{B}_i\right)}{\left(2^{\mathcal{T}_i}-1\right)}-\frac{\sigma_0^2}{\varrho^I_i}\geq 0,\forall i\in K^I,\tag{\ref{eqn:trace_simp_null}a}\label{eqn:trace_simp_nullCa}\\
	&\,\quad \sum_{i\in K^I}\mathrm{tr}\left(\mathbf{B}_i\right)+\mathrm{tr}\left(\mathbf{D}\right)\leq P_\text{max},\tag{\ref{eqn:trace_simp_null}b}\label{eqn:trace_simp_nullCb}\\
	&\,\,\quad \mathbf{B}_i\succeq \mathbf{0},\forall i\in K^I, \mathbf{D} \succeq\mathbf{0},\tag{\ref{eqn:trace_simp_null}c}\label{eqn:trace_simp_nullCc}
\end{align}
where $\mathbf{S}_i^{E,I} = \sum_{t\in K^E}\mathbf{h}^{E,I}_{t,i}\left(\mathbf{h}^{E,I}_{t,i}\right)^H\in\mathbb{C}^{\left(M-K^I+1\right)\times \left(M-K^I+1\right)}$, $\mathbf{S}^E=\sum_{t\in K^E}\mathbf{h}^{E}_{t}\left(\mathbf{h}^{E}_{t}\right)^H\in\mathbb{C}^{\left(M-K^I\right)\times \left(M-K^I\right)}$, $\mathbf{B}_i=P^I_i\mathbf{b}_i\left(\mathbf{b}_i\right)^H\in\mathbb{C}^{\left(M-K^I+1\right)\times \left(M-K^I+1\right)}, \forall i\in K^I$ and $\mathbf{D}=\sum_{j\in K^E}P^E_j\mathbf{d}_j\left(\mathbf{d}_j\right)^H\in\mathbb{C}^{\left(M-K^I\right)\times \left(M-K^I\right)}$.

\begin{Lemma}\label{lemma:1}
	The optimal solution $\mathbf{D}$ of problem $\text{(P2.1)}$ satisfy $\mathbf{D}=\mathbf{0}$.
\end{Lemma}
\begin{IEEEproof}
	Please refer to Appendix \ref{app:A} for detailed proof.
\end{IEEEproof}

\begin{Rem}
	Lemma \ref{lemma:1} demonstrates that the cost of interference elimination in the proposed SWIPT system outweighs the benefits of using dedicated EBs. Therefore, the energy for the EUs is assigned to IB that serves a dual function for both information transmission and power transfer. Additionally, the avoidance of dedicated energy beamforming reduces the system's computational complexity to some extent. It is important to note that Lemma \ref{lemma:1} contradicts the findings in \cite{Xujie}, where $\mathbf{D} \neq \mathbf{0}$ by interference elimination. This discrepancy arises because \cite{Xujie} does not employ a specific interference elimination method and assumes the process to be costless.
\end{Rem}

According to Lemma \ref{lemma:1}, problem $\text{(P2.1)}$ can be further simplified as 
\begin{align}
	&\text{(P2.2): }\max_{\substack{\mathbf{b}_i,\forall i\in K^I}} \quad \sum_{i\in K^I}\mathrm{tr}(\mathbf{S}^{E,I}_{i}\mathbf{B}_i)\label{eqn:trace_simp2}\\
	&\text{s.t. } \frac{\mathrm{tr}\left(\mathbf{h}_i^{I,I}\left(\mathbf{h}_i^{I, I}\right)^H\mathbf{B}_i\right)}{\left(2^{\mathcal{T}_i}-1\right)}-\frac{\sigma_0^2}{\varrho^I_i}\geq 0,\forall i\in K^I,\tag{\ref{eqn:trace_simp2}a}\label{eqn:trace_simp2Ca}\\
	&\,\quad \sum_{i\in K^I}\mathrm{tr}\left(\mathbf{B}_i\right)\leq P_\text{max},\tag{\ref{eqn:trace_simp2}b}\label{eqn:trace_simp2Cb}\\
	&\,\,\quad \mathbf{B}_i\succeq \mathbf{0},\forall i\in K^I,\tag{\ref{eqn:trace_simp2}c}\label{eqn:trace_simp2Cc}
\end{align}

Problem $\text{(P2.2)}$ is also a convex optimization problem, which can be effectively solved by standard optimization method, such as interior point method. The computational complexity of solving problem $\text{(P2.2)}$ is $\mathcal{O}\left(\left(K^I\left(M-K^I+1\right)\right)^{3.5}\right)$. Similarly to problem $\text{(P1.2)}$, the total computational complexity include solving problem $\text{(P2.1)}$ and $K^I$ times eigenvalue decompositions, and the complexity of $K^I$ times null space calculation is $\mathcal{O}\left(K^IM\left(K^I-1\right)^2\right)$. The total computational complexity of null space-based algorithm can be found in Table \ref{table:comput_complxity}. 
% resulting in $\mathcal{O}\left(\left(K^I\frac{\left(M-K^I+1\right)(M-K^I+2)}{2}\right)^{3.5}+K^I(M-K^I+1)^{3}\right)$.

\subsection{Low Computational Complexity Design}\label{sec:lowcomplexity}
In this section, we propose a low computational complexity algorithm for solving the problem $\text{(P2.2)}$. Note that both problems $\text{(P1.2)}$ and $\text{(P2.2)}$ take the contribution of IBs for EUs into consideration, which means EUs can be benefit from the IBs. In practically, transmit power for EUs may much higher than that for IUs \cite{RISS_luo,wqq_IRSWET}. Thus, the benefits from IBs for EUs can be ignored, which motivate us to propose a separate design and achieve a point-to-point communication for IUs.

By utilizing the null space, we eliminate the interference signal, transforming the SINR constraints into signal-to-noise-ratio (SNR) constraints. When considering only the IUs, the optimal IBs can be designed using maximum-ratio transmission (MRT), which maximizes the SNR and can be expressed as
\begin{align}
    &\mathbf{w}_i=P^I_i\frac{\mathbf{N}^I_i\left(\mathbf{N}^I_i\right)^H\mathbf{h}^I_i}{\left|\left|\left(\mathbf{N}^I_i\right)^H\mathbf{h}^I_i\right|\right|},\forall i\in K^I,\label{eqn:bar_w_sub}
\end{align}
where 
\begin{align}
	&P^I_i=\frac{\left(2^{\mathcal{T}_i}-1\right)\sigma_0^2}{\varrho^I_i\beta_i},\nonumber\\
	&\beta_i=\frac{\mathrm{tr}\left(\mathbf{h}^I_i\left(\mathbf{h}^I_i\right)^H\mathbf{N}_i^I\left(\mathbf{N}_i^I\right)^H\mathbf{h}^I_i\left(\mathbf{h}^I_i\right)^H\mathbf{N}_i^I\left(\mathbf{N}_i^I\right)^H\right)}{\left|\left|\left(\mathbf{N}_i^I\right)^H\mathbf{h}^I_i\right|\right|^2},\label{eqn:power_sub}
\end{align}

Take the EUs into consideration, when $i$-th IB is chosen for dual function of information transmission and power transfer, while the other $K^I-1$ IBs are derived from Eq. \eqref{eqn:bar_w_sub}, the optimization problem can be expressed as 
\begin{align}
	&\text{(P3): }\max_{\substack{\mathbf{b}_i}} \quad \mathrm{tr}(\mathbf{S}^{E,I}_{i}\mathbf{B}_i)\label{eqn:trace_simp3}\\
	&\text{s.t. } \frac{\mathrm{tr}\left(\mathbf{h}_i^{I,I}\left(\mathbf{h}_i^{I, I}\right)^H\mathbf{B}_i\right)}{\left(2^{\mathcal{T}_i}-1\right)}-\frac{\sigma_0^2}{\varrho^I_i}\geq 0,\tag{\ref{eqn:trace_simp3}a}\label{eqn:trace_simp3Ca}\\
	&\,\,\,\quad \mathrm{tr}\left(\mathbf{B}_i\right)\leq P_\text{max}-\sum_{l\in K^I,l\neq i}P_l^I,\tag{\ref{eqn:trace_simp3}b}\label{eqn:trace_simp3Cb}\\
	&\,\,\,\quad \mathbf{B}_i\succeq \mathbf{0},\tag{\ref{eqn:trace_simp3}c}\label{eqn:trace_simp3Cc}
\end{align}
where $P_l^I$ is derived from Eq. \eqref{eqn:power_sub}. Problem $\text{(P3)}$ provides the maximal received power when the $i$-th IB is designed for dual function. To maximize the received power, we need to search through the received power values, and the most suitable IB, denoted as $\mathbf{B}_b$, can be determined by
\begin{align}
	\mathbf{B}_b = \arg\max_{t\in K^I} \sum_{i\in K^I,i\neq t}\mathrm{tr}(\mathbf{S}\mathbf{W}_i)+\mathrm{tr}(\mathbf{S}^{E,I}_{t}\mathbf{B}_t),
\end{align}
where $\mathbf{W}_i, \forall i\in K^I,i\neq t$ are derived from Eq. \eqref{eqn:bar_w_sub} and $\mathbf{W}_i=\mathbf{w}_i\left(\mathbf{w}_i\right)^H$, $\mathbf{B}_i$ is obtained from solving problem $\text{(P3)}$. The details of this algorithm are summarized in Algorithm \ref{alg:1}. Note that problem $\text{(P3)}$ is convex optimization problem, which can be effectively solved by standard convex optimization method, such as interior point method. The complexity of solving problem $\text{(P3)}$ is $\mathcal{O}\left(\left(M-K^I+1\right)^{3.5}\right)$, and we need to search through $K^I$ times for the optimal solution. Then, $K^I$ times eigenvalue decompositions and null space calculation are operated to obtain the final beamforming vectors. The total computational complexity of separate design can be found in Table \ref{table:comput_complxity}. 
\begin{algorithm}[t]  
	\small
	\caption{The proposed low computational complexity algorithm.}
	\begin{algorithmic}[1]\label{alg:1}
		\REQUIRE~\ HAP-to-IUs channels $\mathbf{h}^I_i,\forall i\in K^I$, HAP-to-EUs channels $\mathbf{h}^E_j,\forall j\in K^E$, $M$, $\mathcal{T}_i$.
		\ENSURE~\ Optimal solution  $\mathbf{w}^f_i,\forall i\in K^I$.
		\STATE Obtain the information null space $\mathbf{N}^I_i,\forall i\in K^I$ from Eq. \eqref{eqn:NI}.
		\STATE Obtain $\mathbf{w}_i,\forall i\in K^I$ and $P_I^i, \forall i\in K^I$ from Eq. \eqref{eqn:bar_w_sub}-\eqref{eqn:power_sub}.
		\FOR {$i$=1:$K^I$}
			\STATE Obtain optimal $\mathbf{B}_i$ by solving problem $\text{(P3)}$.
			\STATE Obtain the solution $\mathbf{b}_i$ from eigenvalue decomposition of $\mathbf{B}_i$. 
			\STATE Reconstruct $\mathbf{w}_i^b=\sqrt{P_\text{max}-\sum_{l\in K^I,l\neq i} P^I_l} \mathbf{N}^I_i\mathbf{b}_i$, update $\mathbf{w}^f_i=\mathbf{w}^b_i$ and $\mathbf{w}^f_l=\mathbf{w}_l, \forall l\in K^I, l\neq i$.
			\STATE Calculate the total received power $\sum_{t\in K^I,t\neq i}\mathrm{tr}(\mathbf{S}\mathbf{W}_t)+\mathrm{tr}(\mathbf{S}^{E,I}_{i}\mathbf{B}_i)$.
		\ENDFOR
		\STATE Obtain the maximal total received power and the corresponding $\mathbf{w}^f_l$ is the optimal solution.
	\end{algorithmic}
\end{algorithm}

\begin{table*}[t]
	% \small
	\centering
	\caption{The Complexity of Solving Different Problem.}
	\begin{tabular}{m{1.9 cm}<{\centering}|m{9.85 cm}<{\centering}|m{2 cm}<{\centering}|m{2.1 cm}<{\centering}}
	\hline	& \textbf{Computational Complexity} & $K^E=K^I=2$, $M=4$ &$K^E=K^I=16$, $M=64$\\
	\hline
	\textbf{Problem $\text{(P1.2)}$} &  $\mathcal{O}\left(\left((K^I + 1) M\right)^{3.5}+\left(K^I+1\right)M^{3}\right)$ & -& -\\
	\hline
	\textbf{Problem $\text{(P2.2)}$}& $\mathcal{O}\left(\left(K^I\left(M-K^I+1\right)\right)^{3.5}+K^I(M-K^I+1)^{3}+K^I M(K^I-1)^2\right)$& 90.4\% reduction & 68.2\% reduction\\
	\hline
	\textbf{Problem $\text{(P3)}$}& $\mathcal{O}\left(
		K^I\left(\left(M-K^I+1\right)^{3.5}+\left(M-K^I+1\right)^3+M\left(K^I-1\right)^2\right)
		\right)$&97.5 \% reduction& 99.96\% reduction\\
	\hline
	\end{tabular} \label{table:comput_complxity}
\end{table*}

\section{Numerical Results}\label{sec:numericalresult}

In this section, we demonstrate the performance of the proposed null space-based SWIPT system. The signal attenuation at a reference distance of 1 meter (m) is set to 30 dB, and the pathloss exponents for the HAP-EU and HAP-IU channels are set to 2.2 and 3.2, respectively. Additionally, the HAP-to-EU distance is set to 5 m, while the HAP-to-IU distance is set to 50 m. The noise power $\sigma_0^2$ is set to $-84$ dBm, and the Rician factors for both the HAP-EU and HAP-IU channels, i.e., $\kappa_i^I = \kappa_j^E, \forall i \in K^I, j \in K^E$, are set to 5. We assume $K^I = K^E = 2$ and $M = 4$ without loss of generality, with the channel capacity requirements $\mathcal{T}_i,\forall i\in K^I$ set to 8 bps/Hz. 

\subsection{Reduction of Computational Complexity}
Table \ref{table:comput_complxity} presents the computational complexity of solving problems $\text{(P1.2)}$, $\text{(P2.2)}$, and $\text{(P3)}$. For the sake of clarity, we compare the computational complexity using specific values. For the case of $K^E = K^I = 2$ and $M = 4$, the complexity of $\text{(P2.2)}$ is reduced by 90.4\%, and the complexity of $\text{(P3)}$ is reduced by 97.5\%, relative to the original problem $\text{(P1.2)}$. Moreover, for the case of $K^E = K^I = 16$ and $M = 64$, the respective complexity reductions are 68.2\% and 99.96\%, as shown in Table \ref{table:comput_complxity}. These results highlight the suitability of the proposed low-complexity design for scenarios involving a large number of EUs and IUs.

\subsection{Performance with Different Power}
\begin{figure}
	\centering
	\includegraphics[width=0.95\linewidth]{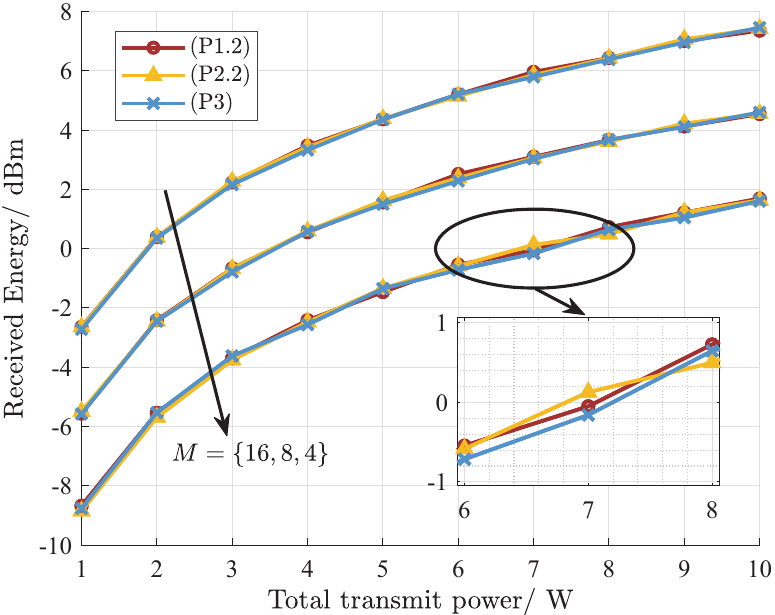}
	\caption{The received power varies with the transmit power.}
	\label{fig:exp1}
\end{figure}

Fig. \ref{fig:exp1} shows the received power with different transmit power $P_{\max}$ and number of transmit antennas $M$. Note that both problem $\text{(P1.2)}$, $\text{(P2.2)}$ and $\text{(P3)}$ exhibit similar trends, wherein the received energy increases with rising transmission power and number of transmit power. Additionally, the results of all three problems demonstrate a high degree of consistency, highlighting the effectiveness of our proposed low-complexity algorithm based on the null space. Notably, although our algorithm disregards the impact of IBs on EUs, and the solution of $\text{(P3)}$ is sub-optimal for original problem $\text{(P2.2)}$, the numerical results indicate that this impact is negligible. This is because the power requirements for communication transmission are significantly lower than those for energy transfer, as mentioned in Section \ref{sec:lowcomplexity}.

\subsection{Performance with Different Communication Requirements and Rician Factors}

\begin{figure}
	\centering
	\includegraphics[width=0.97\linewidth]{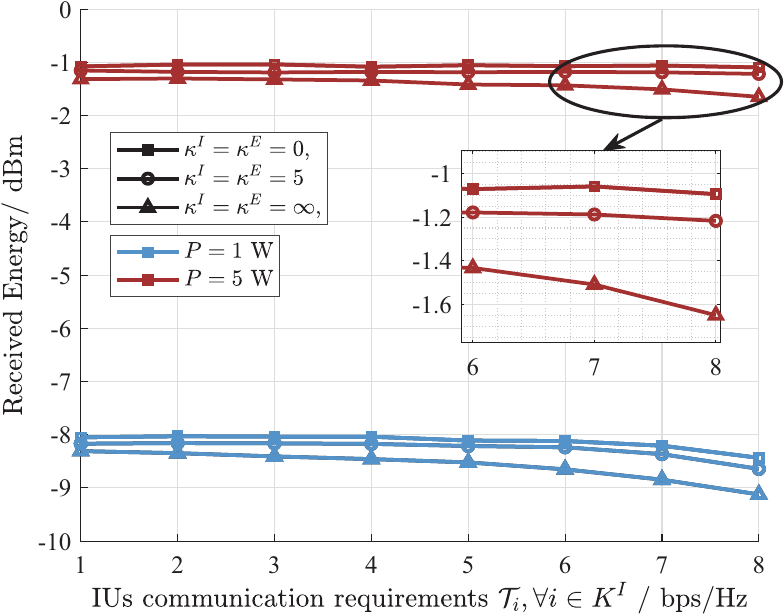}
	\caption{The received power varies with both the communication requirements and the Rician factors.}
	\label{fig:exp2}
\end{figure}

Fig. \ref{fig:exp2} illustrates the variation in received power with different communication requirements, $\mathcal{T}_i, \forall i \in K^I$, and Rician factors. It is observed that the trade-off between WIT and WET results in a decrease in energy performance as communication requirements increase. This effect becomes more pronounced when power resources are limited, e.g., $P = 1$ W. Additionally, higher LoS components, corresponding to higher Rician factors, further reduce energy performance. This is because low-rank channels diminish the spatial multiplexing gain and lead to more severe multiuser interference \cite{WQQ}.

\section{Conclusion}\label{sec:conclusion}
This paper proposed a null space-based SWIPT system capable of eliminating both intra-WIT and inter-WET interference, thereby simplifying the SWIPT architecture. Through the null space design, we demonstrated that dedicated energy beamforming is unnecessary. Additionally, we developed a low-complexity algorithm to address the proposed problem, which further reduces computational complexity. The introduction of this low-complexity algorithm significantly decreases the overall complexity of the null space-based SWIPT system, laying the foundation for SWIPT applications in Massive MIMO scenarios.

% \section{Acknowledgement}
% This work is supported in part by \textcolor{red}{xxxx}.

{\appendices
\section{Proof of Lemma 1}\label{app:A}
Observe from $\text{(P2.1)}$ that $\mathbf{S}_i^{E,i}$ and $\mathbf{S^E}$ can be rewritten as 
\begin{align}
	\mathbf{S}_i^{E,I} &= \sum_{t\in K^E}\mathbf{h}^{E,I}_{t,i}\left(\mathbf{h}^{E,I}_{t,i}\right)^H=\sum_{t\in K^E}\left(\mathbf{N}_{i}^{I}\right)^H\mathbf{h}^{E}_{t}\left(\mathbf{h}^{E}_{t}\right)^H\mathbf{N}_{i}^{I}\nonumber\\
	&=\left(\mathbf{N}_{i}^{I}\right)^H\mathbf{G}\mathbf{N}_{i}^{I},\label{eqn:appReSei}
\end{align}
where $\mathbf{G}=\sum_{t\in K^E}\mathbf{h}_t^E\left(\mathbf{h}_t^E\right)^H$. Similarly, we have $\mathbf{S}^E=\left(\mathbf{N}^E\right)^H\mathbf{G}\mathbf{N}^E$. According to Eq. \eqref{eqn:NI} and Eq. \eqref{eqn:NE}, $\forall \mathbf{m}\in \text{null}\left(\mathbf{H}^I\right)=\mathbf{N}^E$, we have $\mathbf{m}\left[\mathbf{H}^I\right]_{(:, i)}=0, \forall i\in K^I$ and $\mathbf{m}\left[\mathbf{H}_{/i}^I\right]_{(:, j)}=0, \forall i,j \in K^I, i\neq j$. Let $\mathbf{N}^E=\left[\mathbf{n}^E_1,\cdots,\mathbf{n}^E_{M-K^I}\right]$, we can construct $\mathbf{N}^I_i$ by expand $\mathbf{N}^E$ to $\mathbf{N}^I_i=\left[\mathbf{n}^E_1,\cdots,\mathbf{n}^E_{M-K^I}, \mathbf{n}^I_{M-K^I+1}\right]\in\mathbb{C}^{M\times \left(M-K^I+1\right)}$. Thus, $\mathbf{N}^E$ can be represented by 
\begin{align}
	\mathbf{N}^E =\mathbf{N}^I_i \boldsymbol{\Xi} ,
\end{align}
where $\boldsymbol{\Xi}=\left[\mathbf{I}_{(M-K^I)\times (M-K^I)}; \mathbf{0}_{1\times \left(M-K^I\right)}\right]\in\mathbb{C}^{\left(M-K^I+1\right)\times \left(M-K^I\right)}$. Thus, we have 
\begin{align}
	\mathbf{S}^{E}&=\left(\mathbf{N}^{E}\right)^H\mathbf{G}\mathbf{N}^{E}=\boldsymbol{\Xi}^H\left(\mathbf{N}_{i}^{I}\right)^H\mathbf{G}\mathbf{N}_{i}^{I}\boldsymbol{\Xi}\\
	&=\boldsymbol{\Xi}^H\mathbf{S}^{E, I}_i\boldsymbol{\Xi}.
\end{align}
According to Cauchy interlacing theorem \cite{bellman1997introduction}, let $\xi_1^{E}\leq\cdots\leq\xi^{E}_{M-K^I}$ are the eigenvalues of $\mathbf{S}^E$ and $\xi_{1}^{E,I}\leq\cdots\leq\xi^{E,I}_{M-K^I+1}$ are the eigenvalues of $\mathbf{S}^{E,I}_i$, we have 
	\begin{align}
		\xi^{E,I}_j\leq\xi^E_j\leq\xi^{E, I}_{1+j},
	\end{align}
	when $j=M-K^I$, we finally obtain
	\begin{align}
		\xi_{\max}^E=\xi_{M-K^I}^{E}\leq \xi_{M-K^I+1}^{E,I}=\xi_{\max}^{E,I}. \label{eqn:eigvalueEEi}
	\end{align}

Moreover, the Lagrangian of $\text{(P2.1)}$ can be expressed as
\begin{align}
	\mathcal{L}\left\{\mathbf{S}^{E,I}_{i},\mathbf{S}^E,\lambda_i,\beta\right\}=&\beta P_{\text{max}}-\sum_{i\in K^I}\lambda_i\sigma^2_0\nonumber\\
	&+\sum_{i\in K^I}\mathrm{tr}\left\{\boldsymbol{\mathcal{A}}_i\mathbf{B}_i\right\}+\mathrm{tr}\left\{\boldsymbol{\mathcal{C}}\mathbf{D}\right\}
\end{align}

where 
\begin{align}
	&\boldsymbol{\mathcal{A}}_i=\mathbf{S}^{E,I}_{i}+\frac{\lambda_i\mathbf{h}^{I,I}_i\left(\mathbf{h}^{I,I}_i\right)^H}{2^{\mathcal{T}_i}-1}-\beta\mathbf{I},\forall i\in K^I,\\
	&\boldsymbol{\mathcal{C}}=\mathbf{S}_E-\beta\mathbf{I},
\end{align}

Thus, the dual problem of the original problem can be formulated as:
	\begin{align}
		&\min_{\lambda_i\geq0,\beta\geq0}\beta P_{\text{max}}-\sum_{i\in\boldsymbol{\mathcal{K}}^I}\lambda_i\sigma_0^2\label{eqn:D22}\\
		\text{s.t. } \
		&\boldsymbol{\mathcal{C}}\preceq \mathbf{0},\boldsymbol{\mathcal{A}}_i\preceq\mathbf{0},\forall i\in K^I.\tag{\ref{eqn:D22}a}\label{eqn:conD22}
	\end{align}

For the case $\lambda_i=0,\forall i\in K^I$, the constraints \eqref{eqn:conD22} can be simplified as 
\begin{align}
	&\boldsymbol{\mathcal{A}}_i=\mathbf{S}^{E,I}_{i}-\beta\mathbf{I}\preceq\mathbf{0},\forall i\in K^I,\\
	&\boldsymbol{\mathcal{C}}=\mathbf{S}_E-\beta\mathbf{I}\preceq\mathbf{0}.
\end{align}
Thus, we have $\beta \geq \max \left\{\xi_{\max}^E, \xi_{\max}^{E,I}\right\}=\xi_{\max}^{E,I}$ according to Eq. \eqref{eqn:eigvalueEEi}. 
}

For the case at least one $\lambda_i>0,\forall i\in K^I$, we have $\xi\left\{\mathbf{S}^{E,I}_{i}+\frac{\lambda_i\mathbf{h}^{I,I}_i\left(\mathbf{h}^{I,I}_i\right)^H}{2^{\mathcal{T}_i}-1}\right\}\leq\xi\left\{\mathbf{S}^{E,I}_i\right\}+\xi\left\{\frac{\lambda_i\mathbf{h}^{I,I}_i\left(\mathbf{h}^{I,I}_i\right)^H}{2^{\mathcal{T}_i}-1}\right\}$, where $\xi(\cdot)$ denotes the dominant eigenvalue, according to Weyl's inequality \cite{horn2012matrix}. We further have $\xi\left\{\mathbf{S}^{E,I}_i\right\}+\xi\left\{\frac{\lambda_i\mathbf{h}^{I,I}_i\left(\mathbf{h}^{I,I}_i\right)^H}{2^{\mathcal{T}_i}-1}\right\}\geq \xi\left\{\mathbf{S}^{E,I}_i\right\}$ since $\mathbf{h}^{I,I}_i\left(\mathbf{h}^{I,I}_i\right)^H$ is non-negative definite. Thus, for the both cases of $\lambda_i=0,\forall i\in K^I$ and at least one $\lambda_i>0,\forall i\in K^I$, we have $\beta>\xi^E_{\max}$ and $\text{rank}\{\boldsymbol{\mathcal{C}}\}=M-K^I$, which means that $\boldsymbol{\mathcal{C}}$ span the entire space and $\mathbf{D}=\mathbf{0}$. Thus, the proof is completed.
\bibliography{Reference}

\begin{thebibliography}{10}
\providecommand{\url}[1]{#1}
\csname url@samestyle\endcsname
\providecommand{\newblock}{\relax}
\providecommand{\bibinfo}[2]{#2}
\providecommand{\BIBentrySTDinterwordspacing}{\spaceskip=0pt\relax}
\providecommand{\BIBentryALTinterwordstretchfactor}{4}
\providecommand{\BIBentryALTinterwordspacing}{\spaceskip=\fontdimen2\font plus
\BIBentryALTinterwordstretchfactor\fontdimen3\font minus
  \fontdimen4\font\relax}
\providecommand{\BIBforeignlanguage}[2]{{%
\expandafter\ifx\csname l@#1\endcsname\relax
\typeout{** WARNING: IEEEtran.bst: No hyphenation pattern has been}%
\typeout{** loaded for the language `#1'. Using the pattern for}%
\typeout{** the default language instead.}%
\else
\language=\csname l@#1\endcsname
\fi
#2}}
\providecommand{\BIBdecl}{\relax}
\BIBdecl

\bibitem{clerckx2021wireless}
B.~Clerckx, K.~Huang, L.~R. Varshney, S.~Ulukus, and M.-S. Alouini, ``Wireless
  power transfer for future networks: Signal processing, machine learning,
  computing, and sensing,'' \emph{IEEE Journal of Selected Topics in Signal
  Processing}, vol.~15, no.~5, pp. 1060--1094, 2021.

\bibitem{10463057}
K.~Yan, L.~Xiang, and K.~Yang, ``Cooperative target search algorithm for {UAV}
  swarms with limited communication and energy capacity,'' \emph{IEEE
  Communications Letters}, vol.~28, no.~5, pp. 1102--1106, 2024.

\bibitem{lopez2021massive}
O.~L. L{\'o}pez, H.~Alves, R.~D. Souza, S.~Montejo-S{\'a}nchez, E.~M.~G.
  Fern{\'a}ndez, and M.~Latva-Aho, ``Massive wireless energy transfer: Enabling
  sustainable {I}o{T} toward 6{G} era,'' \emph{IEEE Internet of Things
  Journal}, vol.~8, no.~11, pp. 8816--8835, 2021.

\bibitem{wpt_hardware2allocation}
B.~Clerckx, R.~Zhang, R.~Schober, D.~W.~K. Ng, D.~I. Kim, and H.~V. Poor,
  ``Fundamentals of wireless information and power transfer: From {RF} energy
  harvester models to signal and system designs,'' \emph{IEEE Journal on
  Selected Areas in Communications}, vol.~37, no.~1, pp. 4--33, 2019.

\bibitem{massive_WET}
C.~Luo, J.~Hu, L.~Xiang, K.~Yang, and K.-K. Wong, ``Massive wireless energy
  transfer without channel state information via imperfect intelligent
  reflecting surfaces,'' \emph{IEEE Transactions on Vehicular Technology},
  vol.~73, no.~6, pp. 8529--8541, 2024.

\bibitem{SWIPT_hardware}
E.~Boshkovska, D.~W.~K. Ng, N.~Zlatanov, and R.~Schober, ``Practical non-linear
  energy harvesting model and resource allocation for {SWIPT} systems,''
  \emph{IEEE Communications Letters}, vol.~19, no.~12, pp. 2082--2085, 2015.

\bibitem{Hujie_modulation}
J.~Hu, Y.~Zhao, and K.~Yang, ``Modulation and coding design for simultaneous
  wireless information and power transfer,'' \emph{IEEE Communications
  Magazine}, vol.~57, no.~5, pp. 124--130, 2019.

\bibitem{Xujie}
J.~Xu, L.~Liu, and R.~Zhang, ``Multiuser {MISO} beamforming for simultaneous
  wireless information and power transfer,'' \emph{IEEE Transactions on Signal
  Processing}, vol.~62, no.~18, pp. 4798--4810, 2014.

\bibitem{wqq_IRSWET}
Q.~Wu and R.~Zhang, ``Weighted sum power maximization for intelligent
  reflecting surface aided {SWIPT},'' \emph{IEEE Wireless Communications
  Letters}, vol.~9, no.~5, pp. 586--590, 2020.

\bibitem{fluid_SWIPT_LX}
X.~Lin, H.~Yang, Y.~Zhao, J.~Hu, and K.-K. Wong, ``Performance analysis of
  integrated data and energy transfer assisted by fluid antenna systems,'' in
  \emph{ICC 2024-IEEE International Conference on Communications}.\hskip 1em
  plus 0.5em minus 0.4em\relax IEEE, 2024, pp. 2761--2766.

\bibitem{nullspace1}
J.~Tang, A.~Shojaeifard, D.~K.~C. So, K.-K. Wong, and N.~Zhao, ``Energy
  efficiency optimization for {C}o{MP}-{SWIPT} heterogeneous networks,''
  \emph{IEEE Transactions on Communications}, vol.~66, no.~12, pp. 6368--6383,
  2018.

\bibitem{RISS_luo}
C.~Luo, J.~Hu, L.~Xiang, and K.~Yang, ``Reconfigurable intelligent sensing
  surface aided wireless powered communication networks: A
  sensing-then-reflecting approach,'' \emph{IEEE Transactions on
  Communications}, vol.~72, no.~3, pp. 1835--1848, 2024.

\bibitem{WQQ}
Q.~Wu and R.~Zhang, ``Intelligent reflecting surface enhanced wireless network
  via joint active and passive beamforming,'' \emph{IEEE Transactions on
  Wireless Communications}, vol.~18, no.~11, pp. 5394--5409, 2019.

\bibitem{bellman1997introduction}
R.~Bellman, \emph{Introduction to matrix analysis}.\hskip 1em plus 0.5em minus
  0.4em\relax SIAM, 1997.

\bibitem{horn2012matrix}
R.~A. Horn and C.~R. Johnson, \emph{Matrix analysis}.\hskip 1em plus 0.5em
  minus 0.4em\relax Cambridge university press, 2012.

\end{thebibliography}

\end{document}